# High-dimensional Supermode Photonics Enabled by Hierarchical Supersymmetric Transformation

**Yuan Zhong[1], Kaile Chen[1], Qi Lu[1], Chunxue Wang[1], Jingchi Li[1], Yuru Li[2], Zhaohui Li[2,3], Chao Lu[4], Xinchen Ji[1], Yikai Su[1*], and Lu Sun[1*]**

[1]State Key Lab of Photonics and Communications, Department of Electronic Engineering, Shanghai Jiao Tong University, Shanghai 200240, China

[2]School of Microelectronics Science and Technology, Sun Yat-Sen University, Zhuhai 519000, China

[3]Southern Marine Science and Engineering Guangdong Laboratory (Zhuhai), Zhuhai 519000, China

[4]Photonics Research Institute, Department of Electronic and Information Engineering, The Hong Kong Polytechnic University, Hong Kong SAR, China

[*]Corresponding author: Yikai Su and Lu Sun (email: yikaisu@sjtu.edu.cn; sunlu@sjtu.edu.cn)

## Abstract

Modes, a fundamental dimension of light, play a key role in emerging applications such as artificial intelligence (AI), optical computing and quantum information processing. However, conventional multimode waveguides inherently support eigenmodes with non-equidistant effective index distribution, making them susceptible to intermodal crosstalk (CT) caused by phase mismatch, especially when two modes are close in effective index. This limitation constrains the fidelity and scalability of multimode channels in applications such as mode-division multiplexing and mode-encoded high-dimensional entanglement. Supermode photonics, on the other hand, offers a promising route to realizing an equidistant effective index spectrum by geometrically tailoring the coupled waveguide array, thereby enabling the simultaneous control of multiple mode channels and CT suppression for highly parallel and high-dimensional photonic information processing.

Nonetheless, practical deployment of supermode photonics requires precise supermode excitation and detection, which is extremely challenging at the subwavelength scale. Here, we report a hierarchical 2nd-order discrete supersymmetric (DSUSY) transformation method that enables high-purity excitation and extraction of arbitrary target supermodes in a compact and easily scalable architecture. By addressing this long-standing challenge in integrated supermode photonics, our approach provides a new paradigm for harnessing supermodes as a new degree of freedom for encoding, transmitting, and processing information. To prove the feasibility and universality of the method, we experimentally demonstrate the six-supermode multiplexing systems on the silicon-on-insulator and silicon nitride platforms. Leveraging the large effective index spacing between supermodes and the isospectral nature of the DSUSY transformation, the fabricated devices exhibit low insertion losses (< 2.6 dB at the optimal wavelength) and intermodal CT (< -11.1 dB at the optimal wavelength) for all mode channels over a 100-nm wavelength range (1500-1600 nm). The high-speed data transmission experiment performed on the silicon device achieves an aggregate data rate of 1.2 Tbit/s with the bit error rates of all channels below the 7% hard-decision forward error correction threshold, highlighting the potential of supermode photonics for high-capacity on-chip optical communications. More importantly, the proposed DSUSY method can be developed to implement polarization-insensitive architectures, allowing transverse electric and transverse magnetic supermodes to share a common multiplexing structure and therefore enabling compact polarization-supermode hybrid multiplexing. This work lays the foundation for high-dimensional supermode photonics where large numbers of supermode channels are exploited for various applications including but not limited to high-capacity on-chip optical interconnects, highly parallel AI optical computing and high-dimensional quantum information processing.

# Introduction

Driven by the explosive data growth in 5G/6G networks, cloud computing, and artificial intelligence (AI), the need to fully exploit the degrees of freedom of light has become increasingly urgent[1-4]. Modes, an important optical dimension, have attracted sustained research interest and spawned the field of multimode photonics[5-8]. However, conventional multimode waveguides exhibit single-well optical potentials which inherently support eigenmodes with non-equidistantly distributed effective indices ($n_{\mathrm{eff}}$)[9,10]. The close spacing in $n_{\mathrm{eff}}$ makes mode orthogonality highly vulnerable to intermodal crosstalk (CT) arising from the fabrication-induced phase mismatch[7,10], imposing a fundamental obstacle to scalable and high-dimensional mode manipulation[11,12].

Recently, supermode photonics has emerged as a transformative approach to overcome these limitations[13,14]. By tuning the waveguide widths and the inter-waveguide gaps in a coupled waveguide array[15], one can realize a multi-well optical potential that supports supermodes with uniform and maximized $n_{\mathrm{eff}}$ spacing. This property enables the simultaneous and robust control of multiple supermodes, thus preserving low insertion losses (ILs) and minimal CT even in the presence of fabrication imperfections.

However, high-fidelity excitation of a target supermode demands stringent control over the relative amplitudes and phases of the optical fields in individual waveguides, a task that becomes exceptionally challenging at the subwavelength scale. Great efforts have been devoted to exciting supermodes accurately in densely packed straight waveguide arrays[16-18], bent waveguide arrays[19-21], multimode waveguide arrays[22] and multi-core bus waveguides[23]. These architectures deliberately introduce a large phase mismatch between adjacent waveguides, thereby localizing each supermode predominantly within a specific waveguide. Such localization, however, compromises the equidistant $n_{\mathrm{eff}}$ spectrum of the supermodes. More importantly, accessing the supermodes confined to the interior waveguides of the array requires additional mode exchange sections to route them to the outer waveguides, which increases the device footprint and structural complexity and therefore limits its scalability to larger numbers of supermodes.

To address this issue, supersymmetric (SUSY) transformations, originally developed in quantum mechanics[24,25], have been introduced to integrated supermode photonics[26,27]. A wide variety of supermode devices have been developed using SUSY transformations, including single-mode lasers based on waveguide arrays[28-31], graded-index mode converters and multiplexers[32,33] and topological waveguide arrays for the selective excitation of edge and defect states[34-36]. These pioneering works clearly demonstrate the isospectrality of SUSY transformations when engineering the optical potentials of the waveguide arrays[37,38]. However, they mainly focus on the manipulation of a single supermode, whereas the simultaneous control of multiple supermodes, a key requirement for scalable high-dimensional supermode photonics, is still missing.

Here, we propose and experimentally demonstrate a hierarchical 2nd-order discrete SUSY (DSUSY) transformation method that provides a compact and scalable pathway to the high-purity excitation and extraction of an arbitrary number of supermodes. By adiabatically connecting the original waveguide array with its SUSY partner, two target supermodes can be simultaneously accessed via a pair of isolated waveguides. It can be extended to arbitrarily large-scale supermode multiplexing by hierarchically repeating the DSUSY transformations and introducing or eliminating two supermodes at each stage. Compared to our previous work based on the cascade of multiple SUSY structures[39], the hierarchical scheme presented here achieves a larger number of supermode channels within a more compact footprint, improving the scalability and integration density of high-dimensional supermode multiplexing. As proofs of concept, we experimentally demonstrate six-channel supermode multiplexing on both silicon-on-insulator (SOI) and silicon nitride (SiN) platforms. The fabricated devices show low ILs (< 2.6 dB at the optimal wavelength) and intermodal CT (< -11.1 dB at the optimal wavelength) for all channels in the wavelength range of 1500-1600 nm. A high-speed data transmission experiment is also carried out on the silicon device, achieving an aggregate data rate of 1.2 Tbit/s (200 Gbit/s per channel) with the bit error rates (BERs) of all channels below the 7% hard-decision forward error correction (HD-FEC) threshold. Moreover, the proposed method can also be applied to the polarization-insensitive design where transverse electric (TE) and transverse magnetic (TM) supermodes are multiplexed using the same structure. This further simplifies the device architecture and enables compact polarization-supermode hybrid multiplexing. These results lay the foundation for high-dimensional integrated supermode photonics, paving the way towards exploiting supermodes as a new degree of freedom in many applications such as on-chip optical communications[1,3], AI optical computing[40,41] and quantum information technologies[42,43].

# Results

## Effective index engineering and SUSY transformation of supermodes

Figure 1a shows the cross section of a conventional multimode waveguide (left), its corresponding optical potential (middle) and the effective indices ($n_{\mathrm{eff}}$) and mode profiles ($\mathrm{Re}(E_y)$) of the supported TM modes (right). The multimode waveguide can be described by a single square well potential which inherently supports eigenmodes with a non-equidistant $n_{\mathrm{eff}}$ distribution (Fig. 1a, middle). The close $n_{\mathrm{eff}}$ spacing (e. g., that between the blue and green energy levels of the optical potential) will cause severe CT between the corresponding eigenmodes in the presence of fabrication imperfections. To address this issue, we realize a multi-well optical potential by engineering a six-waveguide array, as shown in Fig. 1b. We choose the following parameters for the waveguide array implemented on the SOI platform: $w_1 = w_6 = 411$ nm, $w_2 = w_5 = 497$ nm, $w_3 = w_4 = 550$ nm, $g_1 = g_5 = 222$ nm, $g_2 = g_4 = 124$ nm, and $g_3 = 100$ nm. The design details can be found in Supplementary Note 1. Obviously, the supermode waveguide array exhibits a nearly uniform $n_{\mathrm{eff}}$ spacing of ~0.062 (Fig. 1b, middle) as compared to the multimode waveguide. The equidistant $n_{\mathrm{eff}}$ distribution with maximized spacing will effectively suppress the intermodal CT, a prerequisite for scaling supermode photonics to higher dimensions. However, as one can see from the mode profiles ($\mathrm{Re}(E_y)$) of the TM-polarized supermodes (Fig. 1b, right), high-purity supermode excitation requires precise control over the relative amplitudes and phases of the optical fields in individual waveguides.

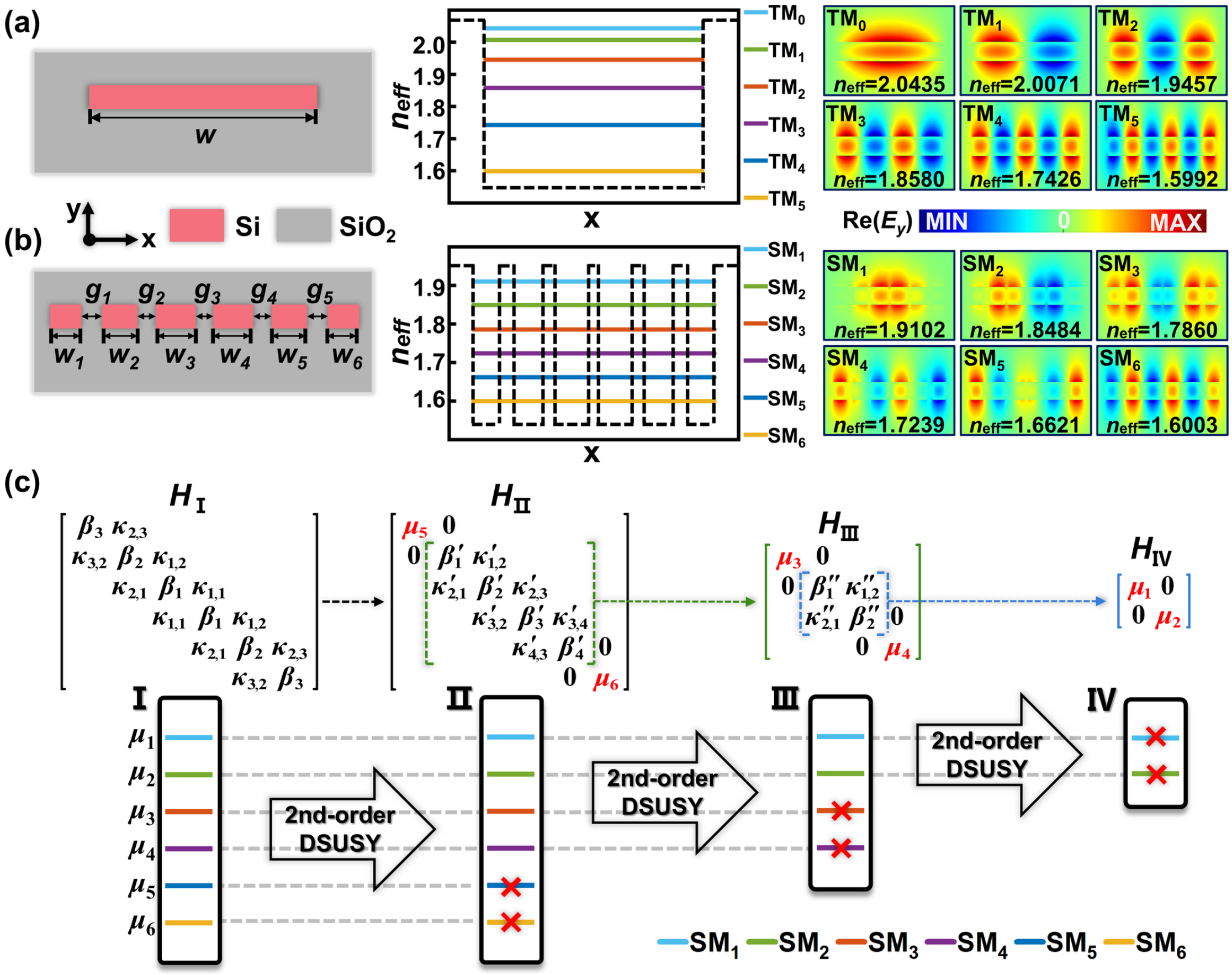

**Fig. 1 | Principle of high-dimensional supermode photonics based on hierarchical SUSY transformation. a,** Cross section of a conventional multimode waveguide (left), its corresponding optical potential and energy levels (middle) and the simulated mode profiles (Re($E_y$)) of the TM modes ($TM_0$-$TM_5$) supported by the waveguide (right). The effective indices of the eigenmodes are non-equidistantly distributed. **b,** Cross section of a six-waveguide array (left), its corresponding optical potential and energy levels (middle) and the simulated mode profiles (Re($E_y$)) of the TM supermodes ($SM_1$-$SM_6$) supported by the array (right). The effective indices of the eigenmodes are equidistantly distributed. **c,** Schematic of the hierarchical 2nd-order DSUSY transformation process. The matrices represent the Hamiltonians $H_{\mathrm{I}}$-$H_{\mathrm{IV}}$ at each DSUSY transformation stage, with the diagonal elements highlighted in red denoting the isolated eigenvalues (top). The spectra of $H_{\mathrm{I}}$-$H_{\mathrm{IV}}$ preserve the isospectrality, with the red crosses indicating the isolation of supermodes at each stage (bottom).

To address this challenge, we propose a hierarchical 2nd-order DSUSY transformation method, as illustrated in Fig. 1c. The original waveguide array can be described by a 6×6 tridiagonal Hamiltonian matrix $H_{\mathrm{I}}$[44], where the diagonal elements represent the propagation constants $\beta_i$ of individual waveguides, and the off-diagonal elements denote the nearest-neighbor coupling coefficients $\kappa_{i,i+1}$ between waveguides. A more detailed derivation is provided in Supplementary Note 2. Then we apply a 2nd-order DSUSY transformation on $H_{\mathrm{I}}$ to acquire the isospectral superpartner matrix $H_{\mathrm{II}}$, with the coupling terms associated with two eigenvalues $\mu_5$ and $\mu_6$ eliminated. It corresponds to a DSUSY-transformed waveguide array with two isolated waveguides. The transformation can be mathematically expressed as

$$H_{\mathrm{original}} - \mu_m I = Q_1 R_1, \tag{1a}$$

$$H_{\mathrm{1st-order\ DSUSY}} = R_1 Q_1 + \mu_m I, \tag{1b}$$

$$H_{\mathrm{1st-order\ DSUSY}} - \mu_n I = Q_2 R_2, \tag{1c}$$

$$H_{\mathrm{2nd-order\ DSUSY}} = R_2 Q_2 + \mu_n I, \tag{1d}$$

where $I$ is the identity matrix, and $Q_{1,2}$ and $R_{1,2}$ are the orthogonal and upper triangular matrices of two QR factorizations, respectively. $H_{\mathrm{2nd-order\ DSUSY}}$ shares the same eigenvalues with $H_{\mathrm{original}}$ and features two isolated diagonal elements $\mu_m$ and $\mu_n$ with vanishing adjacent off-diagonal elements (red matrix elements in Fig. 1c)[39]. Next, we take the 4×4 subblock of $H_{\mathrm{II}}$ with non-vanishing off-diagonal elements (highlighted in the green dashed bracket in Fig. 1c) and apply another 2nd-order DSUSY transformation following Eq. (1) to obtain its SUSY partner $H_{\mathrm{III}}$. $H_{\mathrm{III}}$ has a reduced dimension and another two eigenvalues $\mu_3$ and $\mu_4$ isolated from the rest of the matrix. By hierarchically repeating the 2nd-order DSUSY transformation process, the original matrix can always be decomposed into smaller subblocks with preserved isospectrality. It corresponds to decoupling two supermodes with effective indices $\mu_m$ and $\mu_n$ from the evanescently coupled waveguide array and coupling them into isolated waveguides at each stage. The geometric parameters of the waveguide arrays corresponding to the Hamiltonian matrices $H_{\mathrm{I}}$-$H_{\mathrm{IV}}$ generated during the hierarchical DSUSY transformation process are detailed in Supplementary Note 2.

## Six-supermode multiplexing using hierarchical SUSY transformation

To prove the feasibility of the proposed method in controlling multiple supermodes with an equidistant $n_{\mathrm{eff}}$ distribution, we first design a six-supermode multiplexing device on the SOI platform. As illustrated in Fig. 2a, the device is constructed by reversing the hierarchical DSUSY transformation process and adiabatically connecting the waveguide arrays corresponding to $H_{\mathrm{I}}$-$H_{\mathrm{IV}}$ (located at Points I-IV of the device). The insets above show the cross sections of the waveguide arrays associated with $H_{\mathrm{II}}$-$H_{\mathrm{IV}}$, respectively, whereas that corresponding to $H_{\mathrm{I}}$ is presented in Fig. 1b. The geometry of the adiabatic connection regions can be defined by the following functions:

$$w_i(z) = w_i^{\mathrm{A}} + (w_i^{\mathrm{B}} - w_i^{\mathrm{A}}) f(z), \quad i = 1, \cdots, N, \tag{2a}$$

$$g_j(z) = g_j^{\mathrm{A}} + (g_j^{\mathrm{B}} - g_j^{\mathrm{A}}) f(z), \quad j = 1, \cdots, N-1, \tag{2b}$$

where $w_i(z)$ and $g_j(z)$ are the widths of the $i$-th waveguide and the $j$-th gap as functions of propagation distance $z$, and $w_i^{\mathrm{A,B}}$ and $g_j^{\mathrm{A,B}}$ are those parameters at the starting and end points (denoted by the superscripts A and B, respectively) of the adiabatic connection regions. $f(z)$ can be any slowly varying function satisfying $f(z)\in[0,1]$ and $f(0)=0$, $f(L)=1$, where $L$ is the length of the adiabatic region. In this work, we choose a linear variation function $f(z)=z/L$ for the sake of simplicity. To meet the adiabaticity criterion[45], the lengths between Points I and II, II and III, III and IV are designed to be $L_1$ = 320 μm, $L_2$ = 60 μm and $L_3$ = 50 μm, respectively, yielding a total length of 430 μm for the six-supermode multiplexer. It can be further reduced by employing many advanced techniques, including inverse-design optimization[46], shortcuts to adiabaticity[47], adiabatic passage[48], effective Berry connection minimization[49], and quantum metric engineering[50]. In comparison, our previously reported DSUSY approach requires a total length of ~520 μm to realize a four-supermode multiplexer[39]. Therefore, the hierarchical DSUSY scheme presented here enables more supermode channels within a smaller footprint, offering improved scalability to higher-dimensional supermode multiplexing.

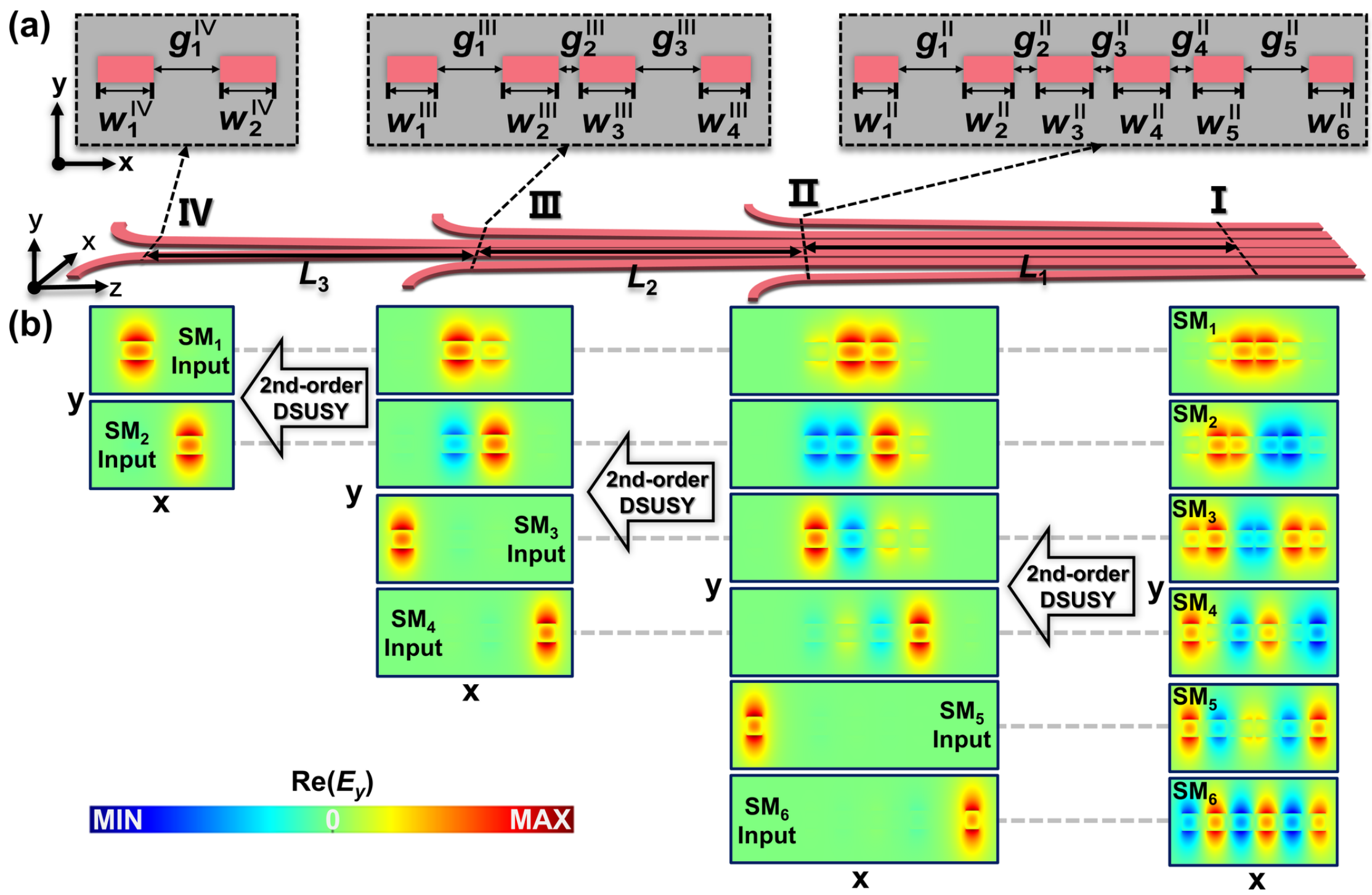

**Fig. 2 | Six-supermode multiplexing enabled by hierarchical SUSY transformation. a,** Six-supermode multiplexer formed by adiabatically connecting the waveguide arrays corresponding to $H_{\mathrm{I}}$-$H_{\mathrm{IV}}$. The cross sections of these waveguide arrays at Points II-IV are shown in the insets. **b,** Simulated mode profiles ($\mathrm{Re}(E_y)$) of the supermodes at Points I-IV, illustrating the adiabatic evolution of these modes and its relationship to the successive 2nd-order DSUSY transformations.

Figure 2b illustrates the evolution of the supermodes by showing the mode profiles ($\mathrm{Re}(E_y)$) along the propagation direction $z$ of the device and its correspondence to the successive 2nd-order DSUSY transformations. The $TM_0$ modes are first launched into two isolated waveguides at Point IV and adiabatically evolve into two supermodes confined to the central waveguides at Point III. Then the $TM_0$ modes are injected into the two outer waveguides at Point III to excite the $SM_3$ and $SM_4$ supermodes, without perturbing the propagation of the SM1 and SM2 supermodes in the inner waveguides. After adiabatic evolution, the $SM_1$-$SM_4$ supermodes become localized to the central four waveguides at Point II, where the signals for the $SM_5$ and $SM_6$ channels are input from the two isolated waveguides. Finally, the six supermodes are adiabatically transformed into the target supermodes supported by the six coupled waveguides at Point I, accomplishing six-supermode multiplexing. The strict isospectrality at each DSUSY transformation stage, together with the adiabatic connection and the maximized $n_{\mathrm{eff}}$ spacing, enable the intermodal CT suppression and therefore the high-purity excitation of high-dimensional supermodes.

To verify the feasibility of the proposed device, we performed three-dimensional finite-difference time-domain (3D FDTD) simulations to track the evolution of the supermodes during propagation. Figure 3a presents the simulated propagation profiles ($\mathrm{Re}(E_y)$) at 1550 nm when the $TM_0$ modes are launched into the respective input ports corresponding to supermodes $SM_1$-$SM_6$. The insets beneath the propagation profiles display the enlarged top views (in the red dashed boxes) and cross-sectional views (in the blue dashed boxes) of the supermode profiles ($\mathrm{Re}(E_y)$) at the output end of the device (Point I). It shows clear evidence that the target supermodes are successfully excited with high purity. The simulated transmission spectra of these supermodes were also extracted to quantitively characterize the performance of the six-supermode multiplexer (Fig. 3b). Over a broad bandwidth of 1500-1600 nm, the ILs are lower than 0.16 dB and the CT values are below -19.4 dB for all six channels (ILs of < 0.07 dB and CT of < -19.8 dB at 1550 nm). These numerical results validate the hierarchical 2nd-order DSUSY transformation as an effective approach to broadband, low-loss and low-CT supermode multiplexing.

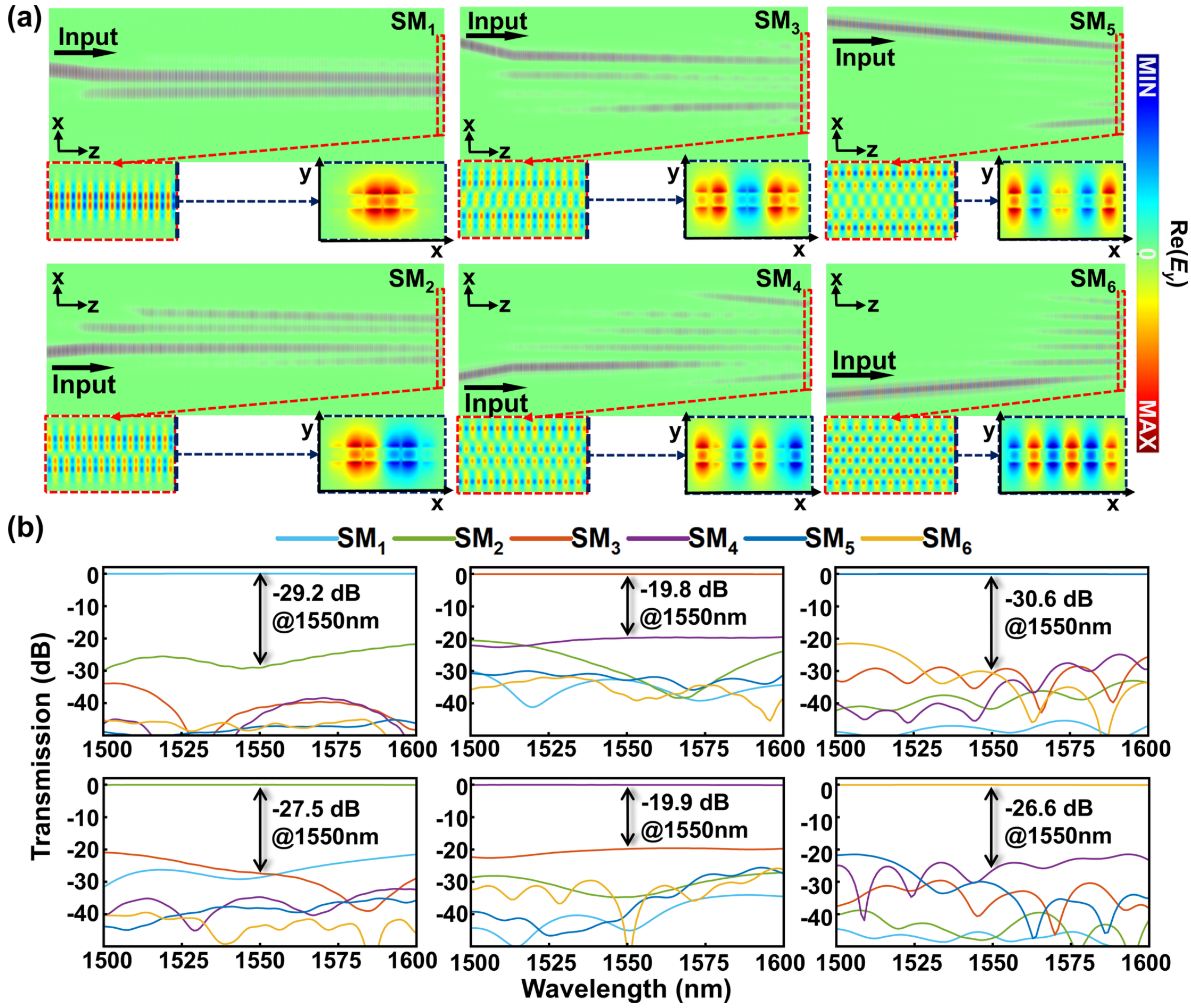

**Fig. 3 | Numerical simulation of the six-supermode multiplexer. a,** Simulated propagation profiles ( $\mathrm{Re}(E_y)$ ) in the adiabatic evolution region when the light is launched from the corresponding isolated waveguides. The insets show the enlarged top-view (in the red dashed boxes) and cross-sectional views (in the blue dashed boxes) of the mode profiles ($\mathrm{Re}(E_y)$) at the output end of the multiplexer. **b,** Simulated transmission spectra of the $SM_1$-$SM_6$ channels over 1500–1600 nm. The double-headed arrows indicate the intermodal CT at 1550 nm.

## Experimental demonstration of the six-supermode (de)multiplexer

The six-supermode (de)multiplexer was fabricated on a SOI wafer comprising a 220-nm-thick silicon layer on top of a 3-µm-thick buried oxide layer using complementary metal-oxide-semiconductor (CMOS)-compatible processes. The device was composed of a pair of mirror-symmetric multiplexing and demultiplexing sections, with the isolated waveguides connected to the grating couplers (GCs) via tapered waveguide bends (Supplementary Note 3). The device patterns were defined by electron-beam lithography (EBL) and transferred onto the top silicon layer by inductively coupled plasma (ICP) dry etching. A 1-µm-thick silica upper cladding layer was deposited on top of the device by plasma enhanced chemical vapor deposition (PECVD). Detailed fabrication procedures can be found in Methods and Supplementary Note 4. Figure 4a shows the optical micrographs of the fabricated device (top) and the zoom-in scanning electron microscope (SEM) images of the waveguide arrays encircled by the green, yellow, blue and red boxes (bottom, labeled as Regions I-IV, respectively).

A tunable continuous-wave (CW) laser and a photodetector (PD) were employed to measure the transmission spectra of all six channels. The experimental setup and the measurement methods are also detailed in Methods and Supplementary Note 4. Figure 4b plots the measured transmission spectra of all supermodes, obtained by launching light into the $SM_1$-$SM_6$ input ports, respectively. The ILs of all channels are kept below 3.5 dB in the wavelength range covering 1500-1600 nm. The best performance is reached at 1544 nm, where all channels exhibit ILs of < 1.8 dB and CT levels of < -11.6 dB. The discrepancies between theory and experiment could be attributed to scattering losses induced by waveguide sidewall roughness and imperfect inter-waveguide coupling arising from the partial filling of the gaps with cladding material[51,52]. The device performance can be further improved through fabrication process optimization. Moreover, the fabrication tolerance analysis reveals that even with substantial deviations in waveguide width (-75 to 250 nm) or gap width (-75 to 75 nm), the (de)multiplexer can maintain low ILs of < 1.65 dB for all channels at 1550 nm. Such a level of robustness is particularly advantageous for large-scale photonic integration (see Supplementary Note 5 for details).

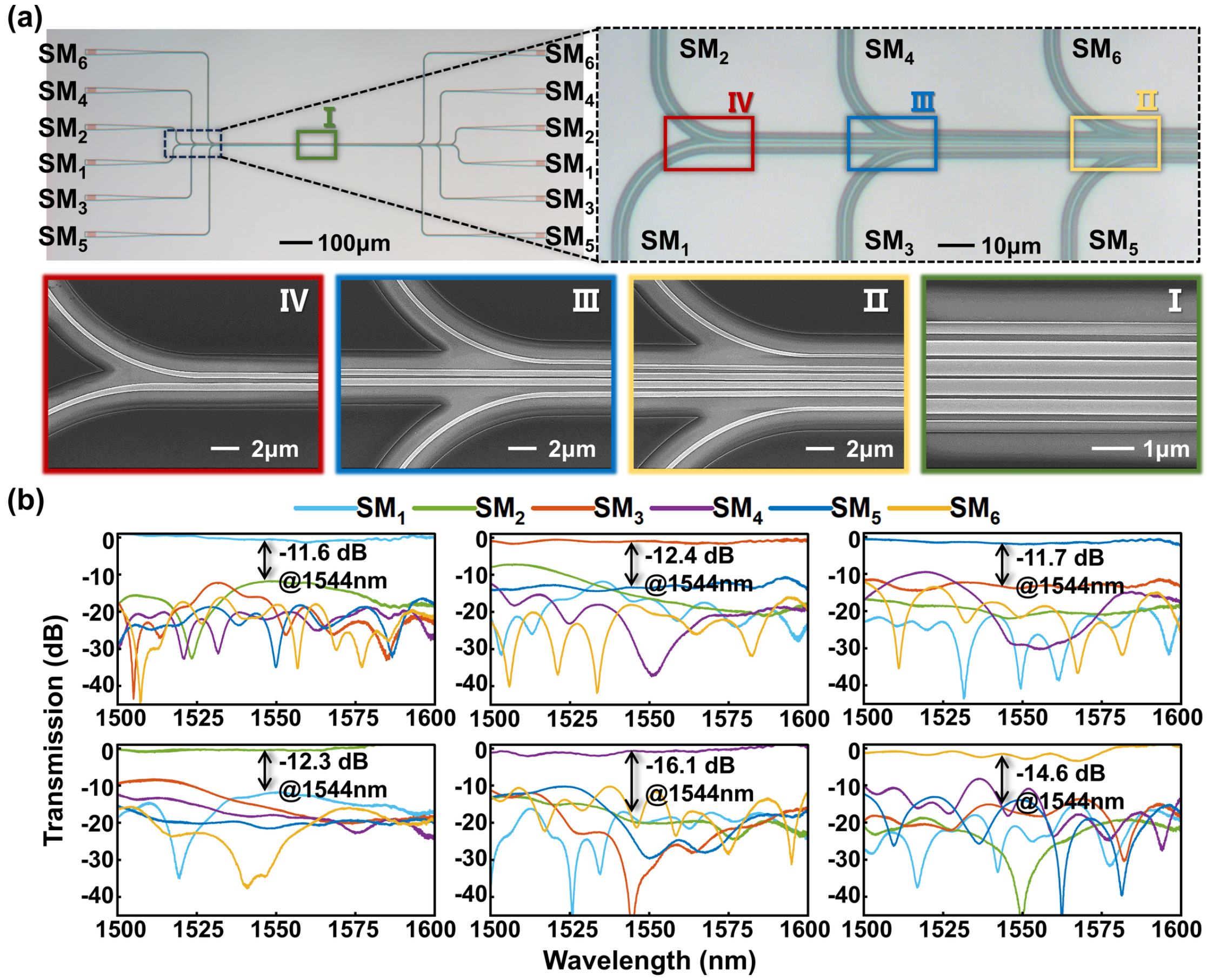

**Fig. 4 | Experimental demonstration of the six-supermode (de)multiplexer. a,** Optical micrograph of the fabricated device (top left), magnified view of the multiplexing region (top right) and zoom-in SEM images of the areas encircled by the green, yellow, blue and red boxes (bottom, labeled as Regions I-IV, respectively). The correspondence between the areas shown in the figures are indicated by the colors of the surrounding boxes and the label numbers. The demultiplexing structure is mirror-symmetric to its multiplexing counterpart. **b,** Measured transmission spectra of the $SM_1$-$SM_6$ channels of the device.

To prove the universality of the proposed hierarchical 2nd-order DSUSY transformation approach, we also implemented a six-supermode multiplexing system on the SiN platform. The design and experimental details are provided in Supplementary Note 6. In simulations, the ILs and CT values are below 0.23 and -18.38 dB in the wavelength range of 1500-1600 nm, respectively (ILs of < 0.17 dB and CT of < -19.5 dB for all channels at 1550 nm). Due to the lower refractive index contrast between the core and cladding materials of the SiN platform, the SiN-based device shows greater fabrication tolerance than its silicon counterpart in the experiment. The measured ILs are less than 3.3 dB from 1505 to 1600 nm for all six channels. The fabricated device achieves the best performance at 1552 nm, with ILs of < 2.6 dB and CT values of < -11.1 dB for all supermode channels. Together, these results extend high-dimensional supermode photonics beyond the silicon platform, highlighting the generality of the hierarchical DSUSY method in supermode multiplexing across diverse integrated photonic platforms.

## High-speed data transmission experiment

To assess the viability of the proposed device for practical applications, a high-speed data transmission experiment was conducted by sequentially launching a high-baud-rate signal into each channel of the silicon-based six-supermode multiplexing system[53,54]. The experimental setup and the digital signal processing (DSP) flowcharts for the transceiver are illustrated in Fig. 5a, b. A Nyquist-shaped 50-GBaud 16-quadrature amplitude modulation (16-QAM) signal at a carrier wavelength of 1544 nm was transmitted through each supermode channel, yielding a data rate of 200 Gbit/s per channel and a capacity of 1.2 Tbit/s in total. At the receiver, offline DSP comprising a multiple-input multiple-output feedforward equalizer (MIMO-FFE) followed by maximum-likelihood sequence detection (MLSD) was applied to compensate for the transmission impairments. More details about the experimental setup and the DSP procedures are provided in Methods and Supplementary Note 7. As shown in Fig. 5c, the BERs of all six supermode channels are below the 7% HD-FEC threshold of $3.8\times10^{-3}$, indicating that error-free transmission can be achieved by adopting standard FEC coding. The recovered 16-QAM constellation diagrams in Fig. 5d show well-resolved symbol clusters for all these channels, validating high-dimensional supermode

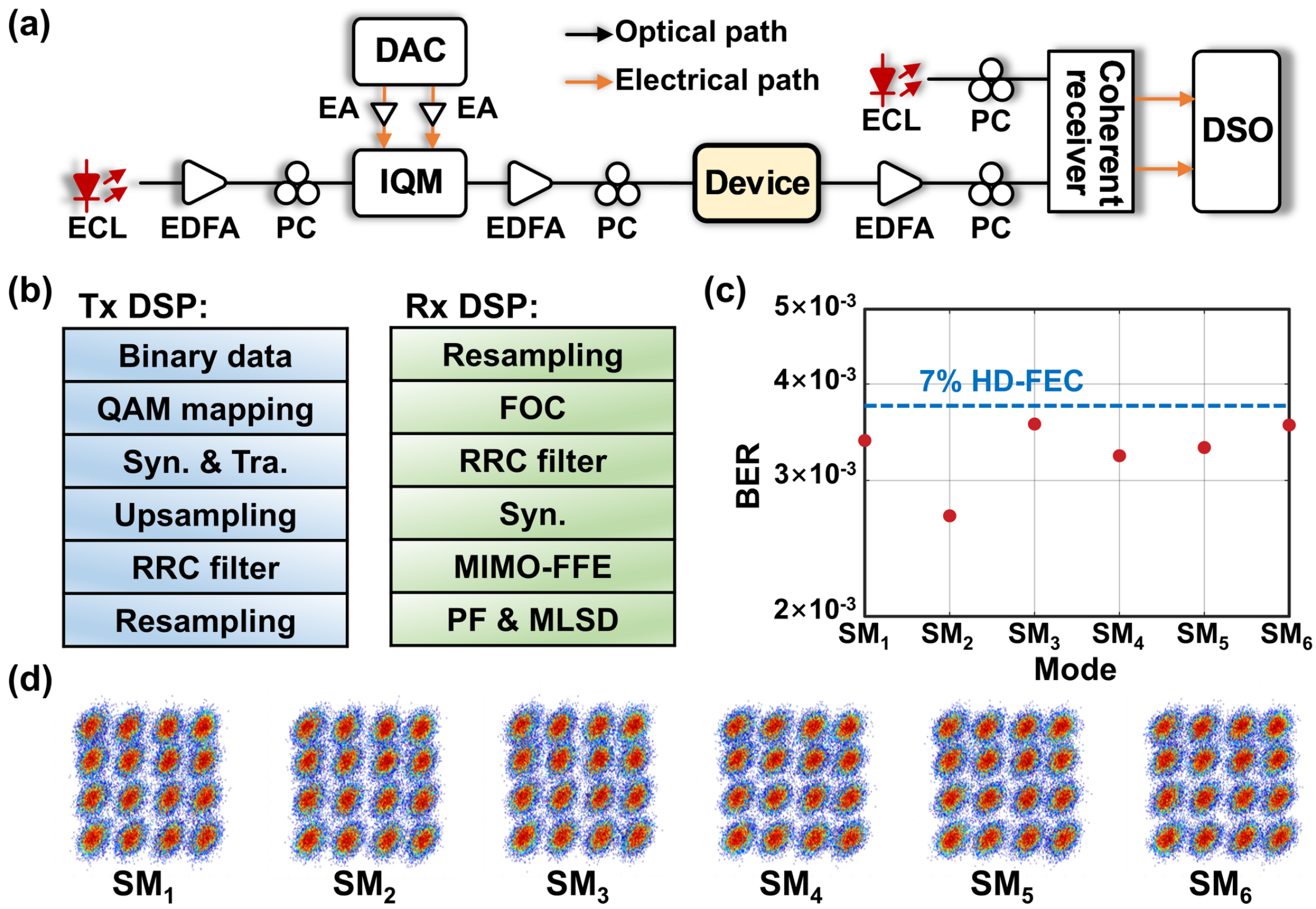

**Fig. 5 | High-speed data transmission experiment based on the six-supermode multiplexing system. a,** Experimental setup for data transmission at 200 Gbit/s per channel. The black and orange lines represent the optical and electrical links, respectively. ECL: external cavity laser, EDFA: erbium-doped fiber amplifier, PC: polarization controller, DAC: digital-to-analog converter, EA: electrical amplifier, IQM: in-phase/quadrature modulator, DSO: digital storage oscilloscope. **b,** DSP flowcharts for the transmitter (Tx) and the receiver (Rx). **c,** Measured BERs of the six supermode channels. The horizontal dashed line denotes the 7% HD-FEC threshold of $3.8\times10^{-3}$. **d,** Recovered 16-QAM constellation diagrams for the $SM_1$-$SM_6$ channels.

photonics as a promising platform for high-capacity on-chip optical interconnects.

More importantly, polarization and supermode multiplexing can be combined to further expand the system capacity. By increasing the thickness of the top silicon layer to 420 nm, we design a four-waveguide array that supports four TE and four TM supermodes with common $n_{\mathrm{eff}}$ spacing on the SOI platform (Supplementary Note 8). This spectral property allows the same device structure, developed based on the hierarchical 2nd-order DSUSY transformation, to simultaneously multiplex and demultiplex the four supermodes of both polarizations. In this way, the as-designed device can support eight channels, arising from the combination of four supermodes and two polarizations. Notably, the total device length remains 320 μm despite the twofold increase in channel number compared to the single-polarization case, avoiding the footprint penalty typically associated with channel scaling. Numerical results prove that the eight channels exhibit ILs of < 0.15 dB and CT of < -17.4 dB in a broad bandwidth spanning from 1500 to 1600 nm (ILs of < 0.052 dB and CT of < -19.3 dB for all channels at 1550 nm), manifesting polarization-supermode hybrid multiplexing as a viable route to doubling the number of accessible channels within a compact footprint. Our vision of the possible applications of polarization-supermode hybrid multiplexing is illustrated in Fig. 6, showcasing its potential beyond high-capacity on-chip optical interconnects. By providing more degrees of freedom for light manipulation, this technique could also enable highly parallel AI optical computing and high-dimensional entanglement for quantum computing, to name just a few.

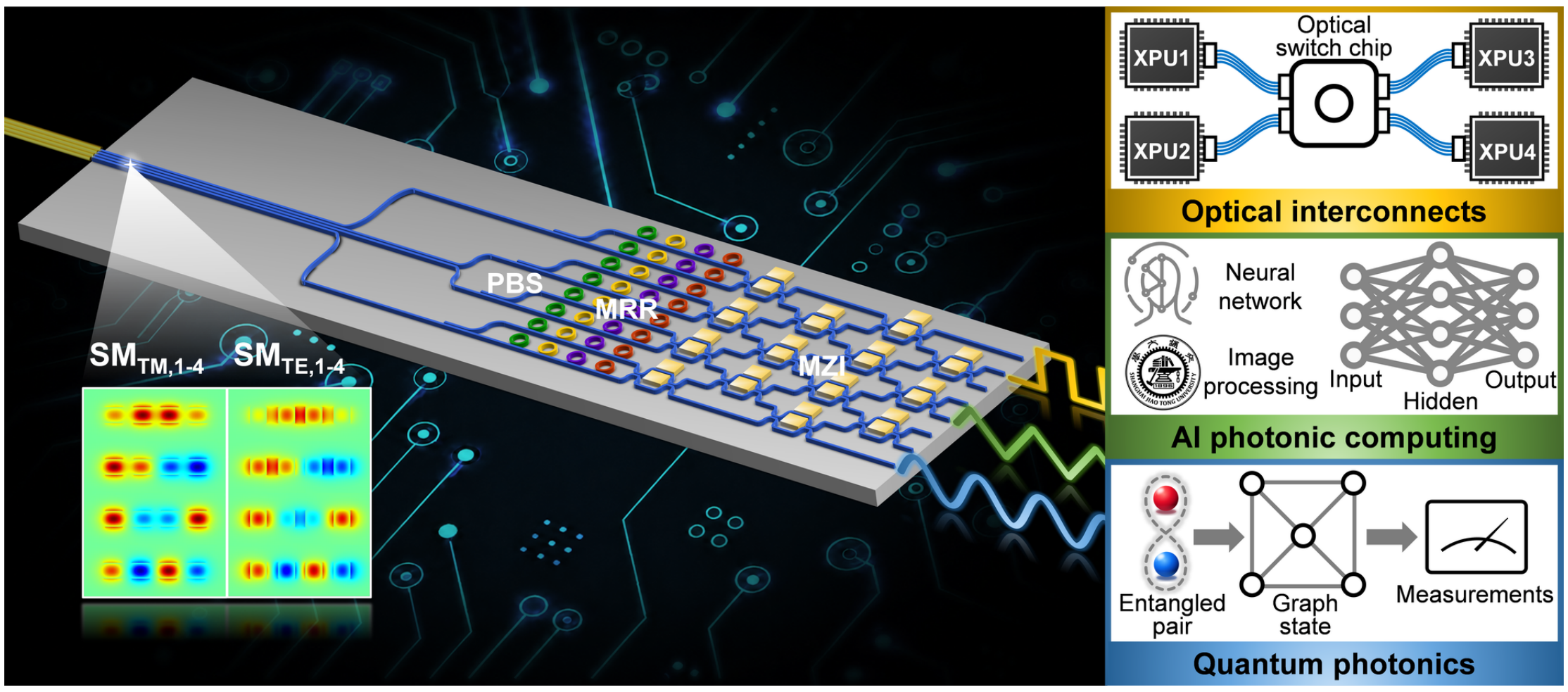

**Fig. 6 | Polarization-supermode hybrid multiplexing and its potential applications.** Combined with programmable photonics, polarization-supermode hybrid multiplexing enables high-speed on-chip optical interconnects between XPUs, highly parallel AI optical computing and high-dimensional quantum entanglement. PBS: polarization beam splitter, MRR: micro-ring resonator, MZI: Mach-Zehnder interferometer, XPU: a generic term for CPUs, GPUs and other processing units.

## Discussion

In conclusion, we have proposed and experimentally demonstrated a hierarchical 2nd-order DSUSY transformation method for high-dimensional supermode photonics. By engineering the multi-well optical potential of a coupled waveguide array, we achieved six supermodes with uniform and maximized $n_{\mathrm{eff}}$ spacing, which effectively suppresses the coupling between orthogonal modes and the resulting increase in intermodal CT under perturbations. To realize high-purity excitation and extraction of these high-dimensional supermodes, a systematic methodology was developed by leveraging the isospectrality of the hierarchical 2nd-order DSUSY transformation and the adiabatic mode evolution[15]. Our work establishes a new paradigm that unlocks the supermode dimension as a new degree of freedom for high-dimensional light manipulation in integrated photonics.

As proof-of-concept demonstrations, we experimentally realized six-channel supermode (de)multiplexers on both SOI and SiN platforms. The fabricated devices show low ILs and intermodal CT for all channels in a broad wavelength range covering 1500-1600 nm (ILs of < 2.6 dB and CT of < -11.1 dB at the optimal wavelength). Compared to our previous work based on DSUSY transformation[39], the hierarchical architecture can achieve more supermode channels with a smaller device footprint of $430 \times 5.6\ \mu\mathrm{m}^2$, demonstrating improved scalability of the proposed method. Furthermore, a high-speed data transmission experiment was also performed on the silicon device by transmitting a 200-Gbit/s 16-QAM signal through each supermode channel. The total capacity of the system reaches 1.2 Tbit/s, with the measured BERs of all six channels falling below the 7% HD-FEC threshold. These results validate the practical viability of the hierarchical DSUSY method for realizing high-capacity on-chip optical communications using high-dimensional supermode multiplexing.

Beyond the present implementation, this work could inspire a wide range of new directions for future studies. For example, the design method can be applied to various integrated photonic platforms such as thin-film lithium niobate[55,56], aluminum gallium arsenide[57,58] and indium phosphide[59,60]. Extending the design to simultaneously support both TE and TM supermodes would enable compact polarization-supermode hybrid multiplexing, which effectively doubles the number of accessible channels without increasing the device footprint (Supplementary Note 8). Besides, low-loss Euler bends have also been designed to route the supermodes on a chip with negligible ILs and intermodal CT (Supplementary Note 9). Most importantly, the system capacity could be further scaled through multidimensional multiplexing, with the total number of channels determined by the product of the available supermodes, polarizations, wavelengths, guided modes supported by each waveguide, etc. Overall, this work provides a scalable and broadly applicable framework for high-dimensional supermode photonics, harnessing supermodes as a new degree of freedom for many signal processing applications, including on-chip optical interconnects[1,3], AI optical computing[40,41] and high-dimensional entanglement-based quantum computing[42,43].

# Methods

## Sample fabrication

The silicon device was fabricated on a SOI wafer with a 220-nm-thick top silicon layer and a 3-μm-thick buried oxide layer. The wafer was first cleaned in ultrasonic baths of acetone and isopropyl alcohol (IPA), followed by $O_2$ plasma treatment. The device patterns were defined by EBL (Vistec EBPG 5200$^+$) and transferred onto the top silicon layer by ICP dry etching (SPTS DRIE-I) with an etching depth of 220 nm. Then the above steps were repeated to fabricate the GCs with an etching depth of 70 nm. Finally, a 1-μm-thick silica upper cladding layer was deposited over the device by PECVD (Oxford Plasmalab System 100). The SiN device was fabricated on a SiN-on-insulator wafer (a 400-nm-thick SiN layer on top of a 3-μm-thick buried oxide layer) following a similar procedure. Since the SiN device employs edge couplers for fiber-to-chip coupling, only a single-step ICP dry etching (NMC GSE200plus) with an etching depth of 400 nm was required. After that, a 2-μm-thick silica upper cladding layer was deposited by PECVD to protect the device. The fabricated devices were inspected using an optical microscope and SEM (Zeiss Ultra Plus). More details can be found in Supplementary Note 4.

**Optical characterization**

The fabricated devices were characterized using a tunable CW laser (Santec TSL-770) and a PD (Santec MPM-210). A fiber PC (Thorlabs FPC032) was used to adjust the polarization of the input light. For the silicon device, the TM-polarized light was coupled into and out of the chip through GCs. For the SiN device, edge couplers were employed for fiber-to-chip coupling, followed by a PBS to improve the polarization selectivity. After calibration using an optical power meter, the laser wavelength was swept from 1500 to 1600 nm, with the output power transmitted through the device recorded by the PD. The measured transmission spectra were then normalized to those of the reference GCs and edge couplers with a PBS. More details can be found in Supplementary Note 4.

**High-speed transmission**

The experimental setup and the offline DSP procedures for the 50-GBaud 16-QAM signal transmission are illustrated in Fig. 5a, b. At the transmitter, the incoming bitstream was first mapped to 16-QAM symbols, followed by upsampling, pulse shaping via a root-raised-cosine (RRC) filter and resampling. The resulting digital signals were converted to analog signals by a 100-GSa/s DAC (Micram DAC4) and amplified by an EA to drive a 35-GHz I/Q modulator. The modulator was biased at the null point for carrier-suppressed modulation. The optical carrier from an ECL with a 15-kHz linewidth and a 10-dBm output power was amplified by an EDFA before being fed into the optical modulator. To compensate for the device ILs, a second EDFA was employed to boost the modulated signals before launching them into the six-channel supermode (de)multiplexer. At the receiver, the transmitted optical signals were pre-amplified, coherently detected and then digitized using a 160-GSa/s DSO (LeCroy 36Zi-A). The offline DSP included resampling, frequency offset estimation, matched filtering, synchronization and MIMO-FFE for linear distortion compensation. Finally, a post filter and MLSD were applied to mitigate the MIMO-induced noise enhancement before BER evaluation. Please see Supplementary Note 7 for more details.

**Author contributions**

Y. Z. and L. S. conceived the idea. Y. Z., K. C., and Q. L. performed the theoretical derivation and numerical simulation. Y. Z. designed and fabricated the devices. Y. Z. and J. L. carried out the experimental measurements. X. J., C. W., and Y. L. assisted the fabrication and characterization. Z. L. and C. L. were involved in the discussions about theory and data analysis. Y. Z. and L. S. wrote the manuscript with input from all authors. L. S. and Y. S. supervised the project.

**Acknowledgements**

The work was supported in part by the National Key Research and Development Program of China (2023YFB2905503(C. L.)) and the National Natural Science Foundation of China (62475146(L. S.) and 62341508(Y. S.)). The authors thank the Center for Advanced Electronic Materials and Devices (AEMD) of Shanghai Jiao Tong University (SJTU) and Hangzhou Brightcore Optoelectronics Co., Ltd. for the support in device fabrication.

**Competing interests**

The authors declare no competing interests.

**Data availability**

The data that support the findings of this study are provided in the Supplementary

Information/Source Data file. Source data are provided with this paper.

## Reference

1 Miller, D. A. B. Device requirements for optical interconnects to silicon chips. *Proc. IEEE* **97**, 1166–1185 (2009).

2 Richardson, D. J., Fini, J. M. & Nelson, L. E. Space-division multiplexing in optical fibres. *Nat. Photonics* **7**, 354–362 (2013).

3 Cheng, Q., Bahadori, M., Glick, M., Rumley, S. & Bergman, K. Recent advances in optical technologies for data centers: A review. *Optica* **5**, 1354–1370 (2018).

4 Bogaerts, W. *et al.* Programmable photonic circuits. *Nature* **586**, 207–216 (2020).

5 Xu, H., Dai, D. & Shi, Y. Silicon integrated nanophotonic devices for on-chip multi-mode interconnects. *Appl. Sci.* **10**, 6365 (2020).

6 Bogaerts, W. & Chrostowski, L. Silicon photonics circuit design: Methods, tools and challenges. *Laser Photon. Rev.* **12**, 1700237 (2018).

7 Luo, L.-W. *et al.* WDM-compatible mode-division multiplexing on a silicon chip. *Nat. Commun.* **5**, 3069 (2014).

8 Stern, B. *et al.* On-chip mode-division multiplexing switch. *Optica* **2**, 530 (2015).

9 Su, Y., Zhang, Y., Qiu, C., Guo, X. & Sun, L. Silicon photonic platform for passive waveguide devices: Materials, fabrication, and applications. *Adv. Mater. Technol.* **5**, 1901153 (2020).

10 Williams, C., Zhang, G., Priti, R., Cowan, G. & Liboiron-Ladouceur, O. Modal crosstalk in silicon photonic multimode interconnects. *Opt. Express* **27**, 27712–27725 (2019).

11 Yu, Y., Sun, C. & Zhang, X. Silicon chip-scale space-division multiplexing: From devices to system. *Sci. China Inf. Sci.* **61**, 080403 (2018).

12 Du, J. *et al.* Mode division multiplexing: From photonic integration to optical fiber transmission. *Chin. Opt. Lett.* **19**, 091301 (2021).

13 Szameit, A. & Nolte, S. Discrete optics in femtosecond-laser-written photonic structures. *J. Phys. B: At. Mol. Opt. Phys.* **43**, 163001 (2010).

14 Longhi, S. Quantum‐optical analogies using photonic structures. *Laser Photon. Rev.* **3**, 243–261 (2009).

15 Miri, M. A., Heinrich, M., El-Ganainy, R. & Christodoulides, D. N. Supersymmetric optical structures. *Phys. Rev. Lett.* **110**, 233902 (2013).

16 Liu, L. Densely packed waveguide array (DPWA) on a silicon chip for mode division multiplexing. *Opt. Express* **23**, 12135–12143 (2015).

17 Chen, K. *et al.* Experimental demonstration of simultaneous mode and polarization-division multiplexing based on silicon densely packed waveguide array. *Opt. Lett.* **40**, 4655–4658 (2015).

18 Cerutti, I., Andriolli, N. & Velha, P. Engineering of closely packed silicon-on-isolator waveguide arrays for mode division multiplexing applications. *J. Opt. Soc. Am. B* **34**, 497–506 (2017).

19 Xu, H. & Shi, Y. Ultra-broadband 16-channel mode division (de)multiplexer utilizing densely packed bent waveguide arrays. *Opt. Lett.* **41**, 4815–4818 (2016).

20 Song, W. *et al.* High-density waveguide superlattices with low crosstalk. *Nat. Commun.* **6**, 7027 (2015).

21 Gatdula, R., Abaslou, S., Lu, M., Stein, A. & Jiang, W. Guiding light in bent waveguide

superlattices with low crosstalk. *Optica* **6**, 585–591 (2019).

22 Xu, H. & Shi, Y. Broadband nine-channel mode-division (de)multiplexer based on densely packed multimode waveguide arrays. *J. Lightwave Technol.* **35**, 4949–4953 (2017).

23 Liu, R., Li, H., Zhu, M., Li, G. & Dai, D. Supermode-division-(de)multiplexing with a multi-core silicon photonic bus waveguide. *J. Lightwave Technol.* **42**, 1–7 (2024).

24 Witten, E. Dynamical breaking of supersymmetry. *Nucl. Phys. B* **188**, 513–554 (1981).

25 Cooper, F., Khare, A. & Sukhatme, U. Supersymmetry and quantum mechanics. *Phys. Rep.* **251**, 267–385 (1995).

26 Chumakov, S. M. & Wolf, K. B. Supersymmetry in Helmholtz optics. *Phys. Lett. A* **193**, 51–53 (1994).

27 Miri, M.-A., Heinrich, M. & Christodoulides, D. SUSY-inspired one-dimensional transformation optics. *Optica* **1**, 89–95 (2014).

28 El-ganainy, R., Ge, L., Khajavikhan, M. & Christodoulides, D. Supersymmetric laser arrays. *Phys. Rev. A* **92**, 033818 (2015).

29 Qiao, X. *et al.* Higher-dimensional supersymmetric microlaser arrays. *Science* **372**, 403–408 (2021).

30 Midya, B. *et al.* Supersymmetric microring laser arrays. *Photonics Res.* **7**, 363–367 (2019).

31 Hokmabadi, M. P., Nye, N. S., El-Ganainy, R., Christodoulides, D. N. & Khajavikhan, M. Supersymmetric laser arrays. *Science* **363**, 623–626 (2019).

32 Heinrich, M. *et al.* Supersymmetric mode converters. *Nat. Commun.* **5**, 3698 (2014).

33 Queraltó, G., Ahufinger, V. & Mompart, J. Mode-division (de)multiplexing using adiabatic passage and supersymmetric waveguides. *Opt. Express* **25**, 27396–27404 (2017).

34 Viedma, D., Queraltó, G., Mompart, J. & Ahufinger, V. High-efficiency topological pumping with discrete supersymmetry transformations. *Opt. Express* **30**, 23531–23543 (2022).

35 Liu, X. *et al.* Perfect excitation of topological states by supersymmetric waveguides. *Phys. Rev. Lett.* **132**, 016601 (2024).

36 Queraltó, G. *et al.* Topological state engineering via supersymmetric transformations. *Commun. Phys.* **3**, 49 (2020).

37 Heinrich, M. *et al.* Observation of supersymmetric scattering in photonic lattices. *Opt. Lett.* **39**, 6130–6133 (2014).

38 Longhi, S. Supersymmetric transparent optical intersections. *Opt. Lett.* **40**, 463–466 (2015).

39 Kaile, C. *et al.* Integrated supermode photonics enabled by supersymmetric transformation. Preprint at https://arxiv.org/abs/2604.18999v1 (2026).

40 Feldmann, J. *et al.* Parallel convolutional processing using an integrated photonic tensor core. *Nature* **589**, 52–58 (2021).

41 Dong, B. *et al.* Higher-dimensional processing using a photonic tensor core with continuous-time data. *Nat. Photonics* **17**, 1–9 (2023).

42 Wang, J., Sciarrino, F., Laing, A. & Thompson, M. G. Integrated photonic quantum technologies. *Nat. Photonics* **14**, 273–284 (2020).

43 Labonté, L. *et al.* Integrated photonics for quantum communications and metrology. *PRX Quantum* **5**, 010101 (2024).

44 Yariv, A. Coupled-mode theory for guided-wave optics. *IEEE J. Quantum Electron.* **9**, 919–933 (1973).

45 Sun, X., Liu, H.-C. & Yariv, A. Adiabaticity criterion and the shortest adiabatic mode transformer in a coupled-waveguide system. *Opt. Lett.* **34**, 280–282 (2009).

46 Liu, X. *et al.* Approaching optimal light evolution at adiabaticity control limit in inverse-designed waveguides. *Phys. Rev. Lett.* **135**, 266601 (2025).

47 Guéry-Odelin, D. *et al.* Shortcuts to adiabaticity: Concepts, methods, and applications. *Rev. Mod. Phys.* **91**, 045001 (2019).

48 Menchon-Enrich, R. *et al.* Spatial adiabatic passage: A review of recent progress. *Rep. Prog. Phys.* **79**, 074401 (2016).

49 Wu, S. *et al.* Approaching the adiabatic infimum of topological pumps on thin-film lithium niobate waveguides. *Nat. Commun.* **15**, 9805 (2024).

50 Song, W. *et al.* Fast topological pumps via quantum metric engineering on photonic chips. *Sci. Adv.* **10**, eadn5028 (2024).

51 Yap, K. P. *et al.* Correlation of scattering loss, sidewall roughness and waveguide width in silicon-on-insulator (SOI) ridge waveguides. *J. Lightwave Technol.* **27**, 3999–4008 (2009).

52 Lin, X., Ma, X. & He, J. Void-filling and loss reduction in PECVD silica waveguide devices using boron–germanium codoped upper cladding. *IEEE Photonics Technol. Lett.* **22**, 1491–1493 (2010).

53 He, Y. *et al.* On-chip metamaterial-enabled high-order mode-division multiplexing. *Adv. Photonics* **5**, 056008–056008 (2023).

54 Huang, H. *et al.* Demonstration of terabit coherent on-chip optical interconnects employing mode-division multiplexing. *Opt. Lett.* **46**, 2292–2295 (2021).

55 Zhu, D. *et al.* Integrated photonics on thin-film lithium niobate. *Adv. Opt. Photonics* **13**, 242–352 (2021).

56 Wang, C. *et al.* Integrated lithium niobate electro-optic modulators operating at CMOS-compatible voltages. *Nature* **562**, 101–104 (2018).

57 Chang, L. *et al.* Ultra-efficient frequency comb generation in AlGaAs-on-insulator microresonators. *Nat. Commun.* **11**, 1331 (2020).

58 Pu, M., Ottaviano, L., Semenova, E. & Yvind, K. Efficient frequency comb generation in AlGaAs-on-insulator. *Optica* **3**, 823–826 (2016).

59 Smit, M., Williams, K. & van der Tol, J. Past, present, and future of InP-based photonic integration. *APL Photonics* **4**, 050901 (2019).

60 Melati, D., Alippi, A. & Melloni, A. Reconfigurable photonic integrated mode (de)multiplexer for SDM fiber transmission. *Opt. Express* **24**, 12625–12634 (2016).